\documentclass[12pt]{article}
\pdfoutput=1
\usepackage{graphicx}
\usepackage{epsfig}
\usepackage{epstopdf}
\DeclareGraphicsExtensions{.pdf,.eps,.png,.jpg,.mps}
\usepackage{caption}
\usepackage{float}
\usepackage{color}

\usepackage{url}
\usepackage[implicit=false]{hyperref}
\usepackage{cite}

\def\lsim{\mathrel{\rlap{\lower3pt\hbox{\hskip0pt$\sim$}}
     \raise1pt\hbox{$<$}}}         
\def\gsim{\mathrel{\rlap{\lower4pt\hbox{\hskip1pt$\sim$}}
     \raise1pt\hbox{$>$}}}         

\usepackage{amsmath}
\usepackage{amsfonts}

\begin{document}
\begin{titlepage}

\centerline{\Large \bf Demystifying Arrow of Time}
\medskip

\centerline{Zura Kakushadze}
\bigskip

\centerline{\em Free University of Tbilisi, School of Physics\footnote{\, Zura Kakushadze, Ph.D. is a Full Professor at Free University of Tbilisi, and a Cofounder and the CEO at Quantigic$^\circledR$. Email: \href{mailto:z.kakushadze@freeuni.edu.ge}{z.kakushadze@freeuni.edu.ge} and \href{mailto:zura@quantigic.com}{zura@quantigic.com}.}}
\centerline{\em \#240 Davit Aghmashenebeli Highway, Tbilisi 0159, Georgia}
\medskip
\centerline{(March 21, 2024)}

\bigskip
\medskip

\begin{abstract}
{}Scientific discussions of the arrow of time often get quite confusing due to highly complex systems they deal with. Popular literature then often coveys messages that tend to get lost in translation. The purpose of this note is to demystify the arrow of time by stripping off the unnecessary complexities and thereby simplifying the discussion. We do this by providing examples that are exactly solvable and make it easy to see the root cause of the apparent ``time-irreversibility". We also discuss ``time-reversal" solutions, where the initial state evolves such that it reaches the state which is the same as the initial state moving backward in time. These solutions are simple enough to be comprehensible to a highschooler. We discuss the arrow of time both in the classical and quantum settings, including in the cosmological context. 
\end{abstract}
\medskip
\bigskip
\bigskip
\bigskip

\bigskip
\bigskip
\bigskip

\end{titlepage}

\newpage
\section{Introduction and Summary}

{}The arrow of time, the apparent flow of time only in one direction, is deemed an unsolved problem. It's of interest to both the general public and physicists, due to its scientific, philosophical, spiritual, etc., implications. Popular literature, when discussing scientific works on this subject, often coveys messages that tend to get lost in translation, so to speak. This is because the issue is nontrivial and scientific discussions often get quite confusing due to highly complex systems they deal with. 

{}In the context of classical thermodynamics, the arrow of time can be attributed to the second law of thermodynamics. The latter deals with a complex notion of entropy (see, e.g., \cite{Landau}) and complex systems composed of a large number of particles (atoms or molecules). Typically, such systems aren't computationally tractable, so one usually resorts to macroscopic (course-grained) approximations. This complexity and approximations muddy the waters in the context of how is it that the underlying fundamental dynamics is perfectly time-reversible, yet the systems one studies theoretically and observes experimentally appear not to be time-reversible.

{}In the context of quantum mechanics the discussion gets even more confusing. It relies on complex concepts of entanglement and decoherence (see, e.g., \cite{Lesovik}, \cite{Neumann}, and references therein). The probabilistic nature of quantum mechanics doesn't help either, in fact, it makes things even more confusing. All this complexity, yet again, muddies the waters in the context of how is it that the underlying unitary quantum evolution is perfectly time-reversible, yet an observable system appears not to be.

{}The purpose of this note is to demystify the arrow of time by stripping off the unnecessary complexities and thereby simplifying the discussion. Our goal here isn't to say something earthshakingly new. Instead, the goal here is to (hopefully) say things in ways that are easy to follow (at least, in some parts). We do this by providing examples that are exactly solvable and make it easy to see that the apparent ``time-irreversibility" is due to very special initial conditions as opposed to some (mysterious or complex) fundamental principle(s). This, in fact, is the main message of this note: it's all about the initial conditions.

{}This note is organized as follows. In Section 2 we discuss classical one-dimensional ideal gases, where the kinematics is exactly solvable. In Section 3 we consider mixing such ideal gases to illustrate that the apparent ``time-irreversibility" is due to special initial conditions. Excepting Subsection 3.3 (which involves entropy), Section 3 should be comprehensible to a highschooler with a basic understanding of kinematics. In Section 3.4 we discuss ``time-reversal" solutions, where the initial state evolves such that after some time it reaches the state which is the same as the initial state moving backward in time. In Section 4 we discuss what happens in the quantum mechanical context and again argue that the apparent ``time-irreversibility" is due to special initial conditions. In Section 5 we discuss special initial conditions we observe in the cosmological context. We briefly conclude in Section 6. 

\newpage
\section{Ideal Gas in One Dimension}

{}In this section we consider a simple model of an ideal gas in one spatial dimension. One-dimensional kinematics is especially simple, which simplifies the discussion of the arrow of time. 

{}Consider a one-dimensional ideal gas consisting of identical ``molecules", which we will treat as point-like particles, all with the same mass $m$. Let us assume that this gas is placed in a one-dimensional box of length $L$. Let us further assume that the collisions between the particles are elastic, i.e., the total kinetic energy is conserved. Furthermore, let us assume that the collisions of the particles with the walls of the box are also elastic, i.e., when a particle bounces off a wall, its kinetic energy is conserved while its momentum is reversed. We will also assume that the speeds of the particles are non-relativistic.

{}Consider an elastic collision of particle A moving to the right with velocity $v$ and particle B moving to the left with velocity $u$. (Velocities are signed quantities, so conventionally choosing our one-dimensional $x$ coordinate axis to run from left to right, $v>0$ and $u<0$.) After the collision, let the velocity of particle A be $v^\prime$ and the velocity of particle B be $u^\prime$. The momentum and kinetic energy are conserved:
\begin{eqnarray}
 &&mv + mu = mv^\prime + m u^\prime\\
 &&\frac{1}{2}~mv^2 + \frac{1}{2}~mu^2 = \frac{1}{2}~mv^{\prime\,2} + \frac{1}{2}~mu^{\prime\,2}
\end{eqnarray}
Because the masses are the same, the solution to the above system of equations is given by:
\begin{eqnarray}
 &&v^\prime = u\\
 &&u^\prime = v
\end{eqnarray}
I.e., the particles simply exchange their velocities: particle A is now moving to the left with velocity $u$ and particle B is now moving to the right with velocity $v$. 

{}In one dimension, since the particles are identical, we can treat the above elastic collision as particle A and particle B simply going through each other (instead of bouncing off each other). This simplifies things greatly. Indeed, if we have a large number $N$ of identical particles in a one-dimensional box, if we treat them as simply going through each other, and if we tag each particle with a label (e.g., number them from 1 to $N$), then we can follow each particle through time. Its velocity never changes except when it bounces off a wall, when its velocity reverses. The system is equivalent to $N$ non-interacting particles in a one-dimensional box.   

{}Before discussing mixing of gases, let us discuss the temperature $T$ of the gas. By the equipartition theorem \cite{Waterston}, \cite{Waterston2}, \cite{Boltzmann}, \cite{Boltzmann2}, it is given by ($k$ is the Boltzmann constant):
\begin{equation}
 \frac{1}{2}~kT = \frac{1}{2}~m\langle v^2 \rangle
\end{equation}
Here $\langle v^2 \rangle$ is the average velocity squared of the $N$ particles in the gas.

\section{Mixing Gases}
\subsection{Uniform Velocities}

{}Let us now consider two adjacent one-dimensional boxes, box A and box B, both of length $L$, separated by a wall (a partition), with box B located to the right of box A. Each box contains $N$ particles described above ($N\gg 1$). The $2N$ particles in box A and box B are indistinguishable in the sense that they are all point-like and have the same mass $m$, and have no other distinguishing intrinsic properties (albeit their velocities are not all the same). At the initial time $t=0$, the $N$ particles in each box are randomly distributed along the length $L$ of their box. For simplicity, let us assume that $N$ is even. Let us further assume that $N/2$ particles in box A have the same velocity $v_A$ ($v_A > 0$), the other $N/2$ particles in box A have the same velocity $-v_A$, and the distribution of positive and negative velocities (i.e., the particles moving to the right and to the left) in box A is random. Similarly, let us assume that $N/2$ particles in box B have the same velocity $v_B$ ($v_B > 0$), the other $N/2$ particles in box B have the same velocity $-v_B$, and the distribution of positive and negative velocities in box B is random. Let us further assume (this is not critical) that the ratio $v_A / v_B$ is some irrational number, $v_A > v_B$, and $v_A/v_B - 1 \lsim 1$.  As above, all scattering is elastic (including off the walls). The temperatures in box A and box B are $T_A=mv_A^2/k$ and $T_B=mv_B^2/k$, so $T_A>T_B$.    

{}Now let us instantaneously remove the wall between box A and box B. Since we can think of the particles as simply going through each other, even though all $2N$ particles are identical, it is convenient to think of them as labeled by A and B, depending on which box they were in at $t=0$. Each particle labeled by A then has velocity $v_A$ or $-v_A$, and each particle labeled by B then has velocity $v_B$ or $-v_B$. After some sufficiently long time $t = t_1 \gg L/v_B$, the gases will be thoroughly mixed, with particles labeled by A and B randomly distributed along the entire length $2L$ of the combined box, with the positive and negative velocities also randomly distributed. Let us refer to the initial state at $t=0$ (when the wall between box A and box B was removed) as state-0, and the state at $t=t_1$ as state-1. In state-1 the average velocity squared is $\langle v^2 \rangle_1 = (v_A^2 + v_B^2)/2$, and, from the macroscopic point of view, the temperature of the mixed gas is $T_1 = (T_A + T_B)/2$.  

{}From the microscopic point of view, the time evolution of the system is exactly solvable, for each individual particle labeled by $a=1,\dots,2N$. I.e., if at time $t=0$ we know the position $x_a(0)$ and the velocity $v_a(0)$ of any given particle labeled by $a$, then we know the position $x_a(t)$ and the velocity $v_a(t) = dx_a(t)/dt$ thereof at any later time $t$: $x_a(t) = f(x_a(0) + v_a(0) t)$, where the periodic ``saw" function $f(x) = 2L - |2L - x|$ for $0\leq x\leq 4L$, and $f(x + 4L) = f(x)$ for all $x\in {\bf R}$. On the other hand, from the macroscopic point of view, if we didn't know the microscopic state, state-1 looks like a completely random state of a gas of $2N$ particles with the temperature $T_1$ in a box of length $2L$. However, this state is anything but ``random". 

{}To see this, consider state-2, which is constructed from state-1 by simultaneously flipping the signs of the velocities of all $2N$ particles in the combined box. Let us start from state-2 and follow the evolution of the particles labeled by A and particles labeled by B. This evolution is nothing but playing the evolution of state-0 into state-1 backwards in time. It then follows that state-2, precisely after time $t_1$, will evolve into state-3, in which the positions of the particles are identical to those in state-0, while all velocities have their signs flipped. I.e., starting from a seemingly ``random" state-2, after time $t_1$, we end up with all particles with velocities $\pm v_A$ neatly placed in what used to be box A (i.e., to the left of the wall between the original boxes A and B), and all particles with velocities $\pm v_B$ neatly placed in what used to be box B. This state-3 is a highly non-random state, just as is state-0. 

{}In fact, state-2 is as non-random as state-0, as is state-1. State-1 and state-2 only look random from a coarse-grained macroscopic description, where one is only concerned with quantities such as temperature and average velocity squared. In the above example, because we chose a one-dimensional gas with identical particles, we can solve the kinematics exactly, i.e., we know the position and velocity of each of the $2N$ particles precisely at all times. So, we can start with a highly non-random state-0 and arrive at an equally non-random state-1, which we can then make look ``random", even though it is anything but random. 

{}The above example makes it clear that fundamentally there is no difference whatsoever between state-0 evolving into state-1, and state-2 evolving into state-3. Without looking into microscopic details, macroscopic observers (humans) will (in all likelihood) erroneously identify state-1 as ``random", all the while not giving much thought to state-0, because to them (humans) such a state is ``natural" (humans separate things into containers all the time). And therein lies the conundrum. If ``natural" is defined as being truly random, then state-0 isn't ``natural" at all, nor is any state it evolves into, including state-1. Nor is state-2. Therefore, there is nothing surprising in state-2 evolving into state-3. 

{}The arrow of time, as thought of by macroscopic observers (humans), in the above example is nothing but the evolution of state-0 into state-1 perceived as ``natural", while the evolution of state-2 into state-3 perceived as ``unnatural". (The latter being analogous to a broken egg evolving back into its unbroken state.) However, this macroscopic perception of the arrow of time in terms of what is ``natural" and ``unnatural" is illusory. What's perceived as ``natural" and ``unnatural" relates to the initial conditions. For humans state-0 is ``natural", but this doesn't mean it's truly random or generic. We can prepare such a state because the initial conditions in which we exist are not in any way random or generic to start with. Simply put, the arrow of time is attributable to the initial conditions -- within the assumptions of the above example, that is (we will come back to this below). However, conceptually, the above example, by stripping off various complexities that muddy the waters, allows us to clearly see that there's no fundamental ``arrow of time": state-0 evolving into state-1 and state-2 evolving into state-3 are on the equal footing, and all four states are equally non-random and non-generic (and equally ``natural" or ``unnatural", if one uses the human perception, which isn't rooted in any fundamental principles).         

\subsection{Nonuniform Velocities}

{}One may dismiss the above example as too simplistic, because we assumed the magnitudes of the velocities in each box to be uniform. We can readily relax this. 

{}Consider box A and box B as above, separated by a wall at $t=0$. Each box still has $N$ particles (all $2N$ particles are indistinguishable). In box A the particles have nonuniform velocities corresponding to the Boltzmann distribution \cite{Boltzmann3}, \cite{Landau} (this is not critical -- see below) with the temperature $T_A$, and in box B the particles have nonuniform velocities corresponding to the Boltzmann distribution with the temperature $T_B$. We can still follow each particle's evolution, even though the velocities are nonuniform, as the particles simply go through each other. As above, it is convenient to think of the particles as labeled by A and B, depending on which box they were in at $t=0$. We then remove the wall between the boxes and let the gases mix. After some sufficiently long time $t = t_1 \gg L/\sqrt{k~\mbox{min}(T_A,T_B)/m}$, the gases will be thoroughly mixed, with particles labeled by A and B randomly distributed along the entire length $2L$ of the combined box. As above, let us refer to the initial state at $t=0$ (when the wall between box A and box B was removed) as state-0, and the state at $t=t_1$ as state-1. Then, as above, state-2 is constructed from state-1 by simultaneously flipping the signs of the velocities of all $2N$ particles in the combined box.   

{}From the macroscopic point of view, if we didn't know the microscopic state, which we can follow precisely through the time evolution of the system (and we can do it for each individual particle), state-1 looks like a completely random state of a gas of $2N$ particles with the temperature $T_1 = (T_A + T_B)/2$ in a box of length $2L$. However, this state is anything but ``random". Indeed, state-2, after time $t_1$, will evolve into state-3, in which the positions of the particles are identical to those in state-0, while all velocities have their signs flipped. So, it makes no difference whether the velocities are uniform or not, because each particle's velocity and position can be solved exactly, so we know the microscopic state exactly and there is nothing to hide behind. The macroscopic misconstruing of state-1 and state-2 being ``random" again is related to the illusory (human) attribution of what's ``natural", which isn't rooted in any fundamental principles. The arrow of time again is attributable to the initial conditions.     

\subsection{Entropy and the Second Law of Thermodynamics}

{}At first, there might appear to be something unsettling about state-2 evolving into state-3 in the context of the second law of thermodynamics. The entropy of state-2 is the same as the entropy of state-1, which in turn is (na{\"i}vely) expected to be greater than the entropy of state-0 (which is the same as the entropy of state-3). Then the evolution of state-2 into state-3 would appear to decrease entropy and violate the second law of thermodynamics. But not everything is as it seems. 

{}The Sackur-Tetrode equation \cite{Sackur}, \cite{Tetrode} for the entropy $S$ of $N$ indistinguishable particles in a one-dimensional box of length $L$ is given by:
\begin{equation}\label{ST1}
 S/kN = \ln(L/\lambda N) + 3/2
\end{equation}
Here $\lambda = h / \sqrt{2\pi mkT}$ is the thermal wavelength, and $h$ is the Planck constant. Therefore, the Sackur-Tetrode entropies $S_A$ and $S_B$ of the gases in box A and box B in the previous subsection at $t=0$ are given by:
\begin{eqnarray}
 &&S_A = kN\ln(L/\lambda_A N) + 3kN/2\\
 &&S_B = kN\ln(L/\lambda_B N) + 3kN/2
\end{eqnarray}
Here  $\lambda_A = h / \sqrt{2\pi mkT_A}$ and  $\lambda_B = h / \sqrt{2\pi mkT_B}$. The total entropy at $t=0$ is given by:
\begin{equation}
 S_0 = S_A + S_B = kN^\prime\ln(L^\prime/\lambda^\prime N^\prime) + 3kN^\prime/2
\end{equation}
Here $N^\prime = 2N$ is the number of particles in both boxes A and B, and $L^\prime = 2L$ is the length of the combined box. Also, $\lambda^\prime = h / \sqrt{2\pi mkT^\prime}$, where
\begin{equation}
 T^\prime = \sqrt{T_A T_B}
\end{equation}
On the other hand, if we apply the Sackur-Tetrode equation to state-1, where we have $N^\prime  = 2N$ particles in the combined box of length $L^\prime = 2L$ with the macroscopic temperature $T_1 = (T_A + T_B) / 2$, we get the following entropy for state-1:
\begin{equation}
 S_1 = kN^\prime\ln(L^\prime/\lambda_1 N^\prime) + 3kN^\prime/2
\end{equation}
Here $\lambda_1 = h / \sqrt{2\pi mkT_1}$. Since $\ln(T_1) > \ln(T^\prime)$, we have $S_1 > S_0$.  And since state-2 entropy $S_2 = S_1$ and state-3 entropy $S_3 = S_0$, the evolution of state-2 into state-3 would appear to violate the second law of thermodynamics. However, this apparent ``violation" is as misconstrued and illusory as the time of arrow in this context.  

{}The issue here is that the Sackur-Tetrode equation (\ref{ST1}) doesn't hold for the one-dimensional system of $N$ particles we discuss above. There are several assumptions that go into the derivation of (\ref{ST1}), including that the system is macroscopic (in the zeroth approximation this translates into the number of particles $N$ being large -- see below), that the $N$ particles are indistinguishable, and that the ergodic hypothesis holds (see below). With these assumptions, one way to arrive at the Sackur-Tetrode equation is to use the definition of the entropy $S = \ln(\Omega)$ via the number of microstates $\Omega$ and compute the latter using the phase space approach. Another, perhaps somewhat simpler but equivalent, approach is to use the relation between the entropy and the partition function $Z$:
\begin{equation}\label{ent}
 S = U/T + k\ln(Z) 
\end{equation}
Here $U$ is the average energy (which in the system above is given by $U=NkT/2$). 

{}For a single quantum particle in a one-dimensional box of length $L$ the partition function $Z_1$ can be computed by summing the Boltzmann weight $\exp(-E_i/kT)$ over all energy levels $E_i$. In the continuum (classical) limit this reduces to a Gaussian integral ($x$ is the spatial coordinate, and $p$ is the momentum):
\begin{equation}
 Z_1 = \frac{1}{h}\int_0^L dx \int_{-\infty}^\infty dp~\exp(-p^2/2mkT) = L/\lambda 
\end{equation}
Here $\lambda = h / \sqrt{2\pi mkT}$ is the thermal wavelength. For a single particle in a $D$-dimensional cube, we have $Z_1 = L^D/\lambda^D$, or, more generally, $Z_1 = V/\lambda^D$, where $V$ is a $D$-dimensional volume. For $N$ indistinguishable particles in $D$ dimensions (in the macroscopic limit), we have the following (approximate) expression for the partition function (the division by $N!$ is due to the $N$ particles being indistinguishable -- however, this is an approximation -- see below):
\begin{equation}
 Z = Z_1^N / N! = V^N / \lambda^N N!
\end{equation}
Taking into account that $U = (D/2)kT$ (by the equipartition theorem -- see above), and using the Stirling approximation $\ln(N!) \approx N\ln(N) - N$ for large $N$, using (\ref{ent}) we arrive at the Sackur-Tetrode equation in $D$ dimensions:
\begin{equation}
 S/kN = \ln(V/\lambda^D N) + (D/2 + 1)
\end{equation}   
This reduces to (\ref{ST1}) for $D=1$. 

{}However, (\ref{ST1}) doesn't apply to the system at hand for several reasons. When deriving (\ref{ST1}), one integrates over a 2-dimensional phase space (in the case of a single particle in $D=1$, or a $2ND$-dimensional phase space in the case of $N$ particles in $D$ dimensions) with the canonical measure (i.e., $dxdp/h$), which implies equal probability of all accessible microstates in the phase space in the long run (this is Boltzmann's ergodic hypothesis \cite{Boltzmann4}). This is irrespective of whether one derives the Sackur-Tetrode equation by counting the number of microstates or by computing the partition function (as above). However, the ergodic hypothesis clearly doesn't apply here. Indeed, due to kinematic constraints in $D=1$, we have a system of non-interacting $N$ particles, each of which remains in the same energy state {\em ad infinitum}. Simply put, there is no ergodicity here and integration over the phase space is inapplicable. Each particle's partition function $Z_{1-\mbox{\scriptsize particle}}$ can simply be set to 1.  (In particular, it is independent of the temperature $T$.)

{}Another assumption that goes into the aforesaid derivation is the indistinguishability of the $N$ particles. The $N$-particle partition function is then approximated as $Z_{N-\mbox{\scriptsize particle}} \approx (Z_{1-\mbox{\scriptsize particle}})^N/N!$. Dividing by $N!$ (approximately) accounts for the indistinguishability of the $N$ particles. This is a justified (and good) approximation when $N$ is large and when the energy of each of the $N$ particles can take a large number $M$ of values ($M\gg N$). Thus, for the argument's sake, when the energy spectrum is discrete, we have 
\begin{equation}
 (Z_{1-\mbox{\scriptsize particle}})^N = \left(\sum_i \exp(-E_i/kT)\right)^N 
\end{equation}   
Then, the leading contributions to $(Z_{1-\mbox{\scriptsize particle}})^N$ come from the terms with all different values of $i$ in the $N$th-power expansion, whose multiplicity is $N!$, hence the division by $N!$ to avoid overcounting of the microstates due to the indistinguishability. However, in our case each particle is stuck in the same energy state, so there are no such terms, and there is no division by $N!$. The $N$-particle partition function $Z_{N-\mbox{\scriptsize particle}}$ can simply be set to 1. Then, using (\ref{ent}), the entropy is simply $S=kN/2$ (for $N$ particles in a one-dimensional box of length $L$).\footnote{\, Cf. \cite{Sobol}. Also, it can be argued that an arbitrary function of only $N$ can be added to the entropy \cite{Jaynes}, so $S=kN/2$ is superfluous and equivalent to $S=0$.} Therefore, mixing the $2N$ particles from box A and box B doesn't change the total entropy. So, the process is reversible, which is why state-2 can evolve into state-3, and there is no violation of the second law of thermodynamics. 

{}So, any ``paradox" or ``conundrum" arises only due to an ``ignorant" approach by a macroscopic observer who decides to ignore microscopic realities and describe the system using inadequate macroscopic variables. In this regard it is instructive to consider the evolution of the system from state-2 to state-3 in the context of the Gibbs paradox (also, mixing paradox) \cite{Gibbs} as discussed in \cite{Jaynes} (entropy is subjective). If the macroscopic observer knows nothing about the microstates in state-2, then there is no reason for said observer to suspect that state-3 is anything special, and said observer will assume that the entropy of state-3 is the same as that of state-2. In order to start thinking otherwise, said observer would have to make measurements and discover that in state-3 the temperatures on the left and right sides of the combined box are different. In fact, to appreciate how non-generic state-3 is, said observer somehow would have to figure out that the $N$ particles on the left side (box A) all have absolute velocities $v_A$, while the $N$ particles on the right side (box B) all have absolute velocities $v_B$, all the while believing that in state-2 the $2N$ particles are indistinguishable. I.e., said observer somehow would have to know that the $2N$ particles are not in fact indistinguishable. Again, any ``paradox" or ``conundrum" arises only due the ``ignorance" of said observer. 

{}In this regard it is worthwhile to note the following. While state-0 is ``natural" from a macroscopic observer's viewpoint (humans separate things into containers all the time), preparing state-2 would appear to be highly ``unnatural" from said viewpoint. In fact, for large $N$ it's not even clear if it would be technologically feasible. On the flipside, theoretically, if in state-0 at time $t=0$ we know the position $x_a(0)$ and the velocity $v_a(0)$ of each particle labeled by $a=1,\dots,2N$, then we know the position $x_a(t)$ and the velocity $v_a(t)$ thereof at any later time $t$ as the system is exactly solvable. However, when $N$ is large, in practice we wouldn't know $x_a(0)$ and $v_a(0)$, instead resorting to a macroscopic description using a few variables (temperature, volume, pressure), which would invariably imply that we would know nothing about the reversibility of the mixing of the gases in box A and box B. Also, let us note that state-3, which is the same as state-0 with all velocities reversed, after a sufficiently long time $t_2$ would evolve into a thoroughly mixed state-4, and state-5 obtained from state-4 by reversing the velocities of all $2N$ particles after time $t_2$ would evolve into state-0.    

{}It is further worth noting that instead of $N$ particles in box A and $N$ particles in box B, we could have started from a simpler system with $N$ particles in box A and empty box B. Upon removal of the wall between box A and box B, the analysis is similar to that above with the same conclusions.

\subsection{``Time-reversal" Solutions}

{}State-2 is obtained from state-1 by reversing all velocities. Similarly, state-3 is obtained from state-0 by reversing all velocities. We can construct state-0 which is invariant under the reversal of all velocities, i.e., state-0 and state-3 are the same. This can be achieved as follows. Let $a=1,\dots,N$ label the particles in box A, and let $N$ be even. Let $x_a(0) = x_{a+1}(0)$ and $v_a(0) = -v_{a+1}(0)$, where $a=1,3,5,\dots,(N-1)$. Let the same be the case for the particles (if there are any) in box B. Then state-0 is invariant under the reversal of all velocities. This then implies that we have what can be viewed as a ``time-reversal" solution. If we start with state-2 (notwithstanding any technological difficulties), after time $t_1$ it evolves into state-3, which is the same as state-0, which after time $t_1$ evolves into state-1, which is the same as state-2 with all velocities reversed. I.e., state-2 after time $2t_1$ evolves into its ``mirror" image, state-1, with all velocities flipped w.r.t. state-2, which is the same as state-2 going backward in time (hence the name ``time-reversal" solutions). In the process, after time $t_1$, the system goes through state-0, which is a ``mirror" image of itself. The entropy is constant during this process. To an uninformed macroscopic observer, armed with only na{\"i}ve macroscopic intuition, such solutions may appear bewildering. However, as discussed above, there is nothing ``paradoxical" about such solutions, and the aforesaid bewilderment is of said macroscopic observer's own making. 
 
\section{Inelastic Collisions, Particle Emissions, etc.}

\subsection{Inelasticity}
{}In the above examples we assumed elastic collisions. In real life collisions of molecules in a gas are mostly inelastic, so the kinetic energy isn't conserved. In such inelastic collisions the kinetic energy can be, e.g., transferred to the internal degrees of freedom, not conserved due to changes in the colliding molecules (chemical reactions), emission of other particles (e.g., photons), etc. Inelasticity quickly introduces complexity and obscures time-reversibility. Here we will not discuss the possible sources of inelasticity exhaustively as this is unnecessary to drive the main point. Instead, we will focus on photon emissions below.

{} Before doing so, however, let us mention that what makes the above examples computationally tractable (in fact, exactly solvable) is that they are one-dimensional. Even if we assume perfectly elastic collisions in three dimensions, it is computationally impossible to track a large number $N$ of molecules, even though each collision is perfectly time-reversible. One then models the system using a few macroscopic variables, which brings about the illusory ``time-irreversibility". However, it is instructive to distinguish between the aforesaid computational complexity and other, perhaps more profound, sources of illusory ``time-irreversibility" we turn to next.  

\subsection{Photon Emissions}
{}If a system evolves deterministically (classical mechanics), then time-reversibility depends on whether the underlying dynamics is time-reversible. This is independent of whether the system is computationally tractable. In quantum mechanics the underlying wave function evolves deterministically (unitary evolution), and the underlying dynamics is time-reversible. However, in terms of the observables, the system evolves probabilistically, which complicates the matters quite a bit. Thus, consider emission of a photon in a collision of two gas molecules. The collision is inelastic. The photon quickly moves away from the collision region and becomes part of the environment (reservoir), i.e., it doesn't affect any subsequent collisions of molecules in the gas (see, e.g., \cite{Lesovik}). Therefore, the gas is no longer a closed system and time-reversibility is lost. 

{}To achieve time-reversibility we would have to reverse the photon emission process; i.e., we would have to consider the process where the photon is absorbed in the inverse collision of the original molecules. If this were a purely classical and deterministic process, things would be much simpler (albeit not necessarily computationally tractable when a large number of collisions is involved). However, in quantum mechanics the photon emission and absorption processes are probabilistic. So, in the time-reversed evolution of the system, which involves irradiating the gas with a large number of photons, when dealing with a large number of collisions and a large number of photons absorbed, 
the probability that the final state is the original microscopic state is infinitesimally small, purely due to the probabilistic nature of the dynamics. Assuming photon emission is the only source of inelasticity, the aforesaid irradiation may very well achieve (approximately) the same macroscopic state (as defined by a small number of macroscopic variables such as temperature) as the original one (i.e., restore the original temperature of the gas), but the probability that the resultant microscopic state will be the same as the original microscopic state is indeed infinitesimally small. 

{}It is important to note that this illusory ``time-irreversibility" once again goes back to the initial conditions. The system at hand (a gas emitting photons into a reservoir) is not a generic, highly symmetrical system. Instead, it's a 
very special, highly asymmetrical open system, which is not in equilibrium with the environment. Therefore, there is nothing surprising about its apparent yet illusory ``time-irreversibility". 

{}The upshot here is that we are dealing with highly non-generic, special initial conditions in systems such as above. E.g., in \cite{Lesovik2} it was argued that to time-reverse the dissipation of a wave packet using an external electromagnetic field, it would take prohibitively long time of the order of the age of the universe or longer. That, in itself, doesn't imply time-irreversibility, be it practically or theoretically.  It took approximately 14 billion years for the universe to produce the observers (humans) who can prepare such highly non-generic, special systems (highly localized wave packets, gas confined to a volume that can be allowed to expand into another volume, etc.). It should come as no surprise that reversing time evolution in highly non-generic systems with special initial conditions can take the age of the universe or longer. (Any bewilderment in this regard is of the observer's own making, due to what can be described as the irrationality of what the observer deems ``natural", just as with the examples of mixing gases we discuss above.) Put differently, the fact that it would take 14 billion years or longer to restore a broken egg into its original state is no more surprising than the fact that it took 14 billion years of evolution of the universe to create an egg (not to mention a human to observe its breakage) in the first place.

\subsection{Quantum Wave Packets}
{}In this regard it's instructive to consider a quantum mechanical wave packet inside a one-dimensional box of length $L$ (with impenetrable walls at $x=0$ and $x=L$; i.e., the potential $V(x) = 0$ for $x\in (0,L)$, and is infinite at $x = 0$ and $x = L$). Let the wave packet wave function at $t=0$ be $\Psi(x, 0)$, which we will assume to be a real-valued, smooth, twice-differentiable function for $x \in [0,L]$, and which is nonzero inside the interval $[x_1, x_2]$, $0<x_1<x_2<L$, and vanishes outside said interval. The energy levels $E_n$ and eigenfunctions $\psi_n(x)$ (which satisfy $\psi_n(0) = \psi_n(L) = 0$, $n=1,2,\dots$) of the Hamiltonian ${\widehat H}$ for a particle in the box are given by:
\begin{eqnarray}
 &&{\widehat H}\psi_n(x) = E_n\psi_n(x)\\
 &&E_n = \frac{\pi^2\hbar^2n^2}{2mL^2}\\
 &&\psi_n(x) = \sqrt{\frac{2}{L}}~\sin\left(\frac{\pi n x}{L}\right)\\
 &&{\widehat H} = -\frac{\hbar^2}{2m}~\frac{\partial^2}{\partial x^2},~~~0 < x < L
\end{eqnarray} 
The wave packet at $t=0$ is a superposition of the infinite tower of these eigenfunctions:
\begin{eqnarray}
 &&\Psi(x, 0) = \sum_{n=1}^\infty c_n~\psi_n(x)\\
 && c_n = \int_{x_1}^{x_2} dx~\Psi(x, 0)~\psi_n(x)
\end{eqnarray} 
With our assumption that $\Psi(x, 0)$ is real (which is not critical here), the coefficients $c_n$ then are also real. However, allowing complex phases doesn't change any conclusions below. 

{}The evolution of the wave function is given by the Schr{\"o}dinger equation:
\begin{eqnarray}
 &&i\hbar~\frac{\partial \Psi(x, t)}{\partial t} = {\widehat H}\Psi(x, t)\\
 &&\Psi(x, t) = \exp(-i{\widehat H}t/\hbar)~\Psi(x, 0) = \sum_{n=1}^\infty c_n~\psi_n(x)~\exp(-iE_nt/\hbar)
\end{eqnarray}
After a substantially long enough time $t_1$, the wave function $\Psi(x, t_1)$ is no longer localized. To a macroscopic observer who is only interested in its spatial shape, this wave function will not appear any more special than any random wave function. (This is analogous to our discussion of mixing gases above.) However, $\Psi(x, t_1)$, into which $\Psi(x, 0)$ evolves, is just as special and non-generic as $\Psi(x, 0)$. The wave function $\Psi(x, -t_1)$, which evolves into $\Psi(x, 0)$ from $t=-t_1$ to $t=0$, is just as special as $\Psi(x, 0)$, even though to the aforesaid macroscopic observer this wave function will not appear any more special than any random wave function either. Indeed, $\Psi(x, -t_1)$ is simply the complex conjugate of  $\Psi(x, t_1)$: $\Psi(x, -t_1) = \Psi^*(x, t_1)$. However, $\Psi(x, t)$, for all times $t$, before and after $t=0$, is special and non-generic as it corresponds to a localized wave packet at $t=0$. For the system to evolve into this state, its past state at, say, $t=-t_1$, must be highly special and non-generic. So, we are back to special initial conditions. Any difficulty in time-reversing the dissipation of the wave packet after $t=0$ is due to these special initial conditions, not illusory ``time-irreversibility". 

\section{Special Initial Conditions: Why, How, \dots?}

{}The Trillion Dollar Question then is, how come the initial conditions we observe are so special that we observe the apparent arrow of time? A simple -- and perfectly valid -- answer is the anthropic principle \cite{Dicke}, \cite{Weinberg}. If the initial conditions weren't special, we wouldn't exist to ponder the question.  

{}However, one can wonder if such special initial conditions can arise ``naturally", with the ever-present caveat that what's deemed ``natural" can be subjective and influenced by the very nature of the observers (humans) posing said question. The standard lore, in the cosmological context, goes as follows (see, e.g., \cite{Ashtekar} and references therein). Within the cosmic inflationary scenario, the starting point is the Bunch-Davies \cite{Bunch} vacuum state for the quantum fields, which is homogeneous and isotropic. It is then (implicitly) assumed that at the end of inflation the system can be described using classical evolution with a distribution of states in the classical phase space (position and momentum). This emergence of classical behavior in the universe, which is assumed to start in an intrinsically quantum state, is referred to as ``quantum-to-classical transition" or ``classicalization". However, despite decades of effort (see references in, e.g., \cite{Ashtekar} and \cite{Berjon}), this classicalization is still not completely understood (see, e.g., \cite{Berjon}). Perhaps even more importantly (see below), it is unclear how the initial symmetry of the quantum vacuum state (which is homogeneous and isotropic) is broken such that primordial inhomogeneities and anisotropies arise, which are then expected to produce the large-scale structure we observe in the universe (starts, galaxies, galaxy clusters, etc.). In particular, in the aforesaid standard lore, there doesn't appear to exist a satisfactory derivation as it relates to the evolution of an intrinsically quantum state (Bunch-Davies vacuum) into a classical state with a distribution in the classical phase space (which is then tied to the aforesaid primordial inhomogeneities and anisotropies). Instead, it appears to be an {\em ad hoc} assumption with apparent gaps in logic and math (including in the context of ``squeezed states"; see, e.g., \cite{Martin} and references therein). Therefore, the appearance of primordial inhomogeneities and anisotropies required for the formation of the observed large-scale structure in the universe by no means appears to be settled and is an open question (see, e.g. \cite{Berjon}).

\subsection{Macroscopic Objects}

{}The discussion in the cosmological context of classicalization and primordial asymmetries, however, appears to be unnecessarily confusing by mixing various issues that can be understood much more simply. The universe we observe is certainly not ``classical" but quantum. It's just that quantum effects are minuscule on macroscopic scales, which is why we only observe them at microscopic scales. What follows as it relates to quantum effects is nothing new, but perhaps (hopefully) the way it's stated in the context of how special conditions can arise in the cosmological context is helpful in demystifying this topic.

{}Smallness of quantum effects can be thought of in many ways. A standard way is to look at Heisenberg's uncertainty principle:
\begin{equation}
 \Delta x~\Delta p \geq \hbar / 2
\end{equation}         
Let $M$ be the mass of the object the uncertainties $\Delta x$ and $\Delta p$ in whose position $x$ and momentum $p$ we wish to understand. Let $\Delta v$ be the uncertainty in the velocity $v = p/M$. Then we have
\begin{equation}\label{uncertain}
 \Delta x~\Delta v \geq \hbar / 2M
\end{equation}  
For macroscopic objects the mass $M$ is large (compared with atomic scales) and the r.h.s. of (\ref{uncertain}) is so small that the lower bound it imposes on $\Delta x~\Delta v$ is immaterial in the context of realistic measurement errors $(\Delta x)_m$ and $(\Delta v)_m$ by a macroscopic observer, i.e., $(\Delta x)_m~(\Delta v)_m\gg \hbar/2M$. So, the ``quantumness" of the system is well beyond the measurement precision capabilities of the observer.

{}Another, quintessentially quantum, way of thinking about smallness of quantum effects is to consider quantum tunneling. Let's consider a gas of $N$ particles in a one-dimensional box. Let one of the walls of the box not be impenetrable, i.e., let's have a potential barrier of a finite height at said wall. Let the probability for a single particle to tunnel through the wall be $P$. We don't have to assume that $P$ is exponentially small, it can be, say, $P\sim 0.1$. Assuming the $N$ particles are non-interacting and their number is large, the probability of $M$ particles tunneling though the wall is $\sim P^M$. This probability for any macroscopic subset of $M$ particles to tunnel through the wall (even if $M\ll N$; recall that Avogadro's constant $N_A \approx 6\times 10^{23}$ per mole) is tiny and doesn't affect the macroscopic characteristics of the gas (as measured by a macroscopic observer with realistic measurement precision capabilities). 

{}I.e., macroscopic objects behave classically not because the world around us is classical, and not due to some ``classicalization" scheme, but because quantum effects for large objects are small. So, in the cosmological context what needs to be explained isn't some ``classicalization" scheme, but clumping of matter in the early universe that leads to the formation of macroscopic objects. 

\subsection{Clumping of Matter}

{}However, if this clumping itself is homogeneous and isotropic, the resulting large-scale structure will also be homogeneous and isotropic at all scales, which is not what we observe in the universe. The aforesaid assumption (in the context of the aforesaid standard lore) that quantum fluctuations somehow will break the initial symmetry lacks justification and it is unclear what the mechanism would be for such symmetry breaking. We can readily understand this already in the context of non-relativistic quantum mechanics. Here we could consider a simpler one-dimensional analog of the system we describe below. However, the three-dimensional example is just as easy to follow and is more ``realistic". 

{}Thus, consider an infinite three-dimensional cubic lattice. (The choice of the lattice is not critical here, so long as we have periodicity.) The nodes of the lattice have the coordinates ${\vec r}_{\vec n} = {\vec n} L$, where ${\vec n} = (n_1, n_2, n_3)$, $n_i \in {\bf Z}$ ($i=1,2,3$) are integers, and $L$ is the lattice spacing. Let us now consider a system where we have a particle of mass $m$ (initially -- see below) localized around each node (so we have an infinite number of such particles). Let us label these particles by $a$, and their coordinates and momenta by ${\vec r}_a$ and ${\vec p}_a$. Let us assume that the particles are non-relativistic. Let us include Newtonian gravitational interactions between the particles. Then the Schr{\"o}dinger equation and quantum Hamiltonian are given by:\footnote{\, Here we won't include the effects of wave function collapse due to gravitational self-interaction \cite{Diosi}, \cite{Penrose}.}
\begin{eqnarray} 
 &&i\hbar~\frac{\partial \Psi({\vec r}_a, t)}{\partial t} = {\widehat H}\Psi({\vec r}_a, t)\\
 &&{\widehat H} = -\frac{\hbar^2}{2m}\sum_{a} \frac{\partial^2}{\partial {\vec r}_a^2} - \sum_{a\neq b} \frac{G_Nm^2}{{|{\vec r}_a - {\vec r_b}|}}\label{Ham}
\end{eqnarray}                
Here $G_N$ is Newton's gravitational constant. (Here we assume that the space is static; we will discuss expanding space in a moment.) So, we have unitary evolution with a Hamiltonian that is invariant under translations and rotations. In particular, it preserves all the symmetries of the cubic lattice. 

{}Now let us assume that at $t=0$ the wave function $\Psi({\vec r}_a, 0)$ corresponds to the particles localized around the nodes of the lattice. Let $f(r)$ be a smooth, twice-differentiable function of $r$ ($r\geq 0$), such that $f(r) = 0$ for $r \geq R$, where $R\ll L$. Let ${\vec h}$ be a one-to-one map between the set $\{a\}$ and the set $\{{\vec n}\}$, i.e., for each value of $a$, ${\vec h}(a)$ is a vector ${\vec n}_a\in \{{\vec n}\}$, and the inverse $g$ of ${\vec h}$ gives $g({\vec n}_a) = a$. Let us now take the initial wave function to be
\begin{equation}
 \Psi({\vec r}_a, 0) = \sum_a f(|{\vec r}_a - {\vec n}_a L|)
\end{equation}   
I.e., at $t = 0$ we have a spherically symmetric wave packet localized at each of the nodes of the square lattice. (If so desired, we can regularize the infinite sum by considering the space to be a large 3-torus instead.) Since the Hamiltonian is invariant under lattice shifts, and so is $\Psi({\vec r}_a, 0)$, then $\Psi({\vec r}_a, t)$ is also invariant under lattice shifts at all subsequent times $t>0$. 

{}If the wave packets have random nonzero group velocities, then the lattice symmetry is broken and clumping of wave packets will occur due to the gravitational attraction. The resultant wave function will no longer have the symmetry of the lattice. However, this is not due to ``quantum fluctuations" but the initial conditions. 

{}Generally, in the cosmological context, to have inhomogeneities and anisotropies in the large-scale structure of the universe, there need to be primordial inhomogeneities and anisotropies in the early universe as part of the initial conditions. So, if the universe starts in an intrinsically quantum state, the initial conditions must include inhomogeneities and anisotropies as unitary quantum evolution will not introduce such asymmetries. Otherwise, the universe would be homogeneous and isotropic and/or there would be large quantum probabilities of observing identical macroscopic objects in different directions and/or parts of space. The latter issue is related to the ``classicalization" problem. However, there is no issue if the initial conditions intrinsically include inhomogeneities and anisotropies, which then can result in clumping of matter due to the gravitational interaction, thereby leading to the formation of the observed large-scale structure with macroscopic objects in it, which appear to be classical (rather than quantum) for the reasons discussed above.

{}Let us note that in the above discussion we assumed that space wasn't expanding. Let us consider an expanding spacetime with the metric (in the units where the speed of light $c = 1$)
\begin{equation}
 ds^2 = -dt^2 + a^2(t) d{\vec r}^2
\end{equation}
Here $a(t)$ is the scale factor, which increases with time $t$. Then the Hamiltonian (\ref{Ham}) is modified by simply replacing ${\vec r}_a$ and ${\vec r}_b$ by $a(t){\vec r}_a$ and $a(t){\vec r}_b$, which doesn't affect the invariance of the Hamiltonian under translations and rotations, so the conclusions discussed above are unmodified. 

\subsection{How Can Initial Conditions That Include Us Arise?}

{}If we accept that the initial conditions intrinsically include inhomogeneities and anisotropies, that may appear to imply that the initial conditions would have to be tailored for humans to exist, for the author to be typing these words, and for the reader to be reading them. At least for some, that could be a big ask. However, not all may be as it might appear. If we assume randomness in primordial inhomogeneities and anisotropies (with any possible non-Gaussian contributions; see, e.g., \cite{Ezquiaga} and references therein), in the infinite three-dimensional space this will result in an infinite number of possible configurations of special initial conditions, which can include those resulting in the patch of space we live in. These asymmetries can be due to fluctuations in spatial density of matter and in the momenta. I.e., so long as we assume randomness (possibly with fat tails), what we observe around us would appear to have to happen {\em somewhere} in the infinite space with probability 1. Using business lingo, this is a ``big idea". 

{}This by itself doesn't answer the question of how such random (albeit possibly fat-tailed) asymmetries arose. If we assume that all interactions and matter in the universe are quantum and subject to unitary evolution (which itself is homogeneous and isotropic), then it is difficult to see how such asymmetries would magically arise starting from a symmetric vacuum configuration. I.e., these asymmetries would have to be inherent in the initial quantum state. One can only speculate why they were present in said state, e.g., one could consider big-bounce cosmology \cite{Poplawski}, ekpyrotic and cyclic models \cite{Lehners}, etc.    

{}Logically, there is another avenue, to wit, to forgo some of the assumptions discussed above. One of these assumptions is unitary evolution of quantum systems. Another is that all interactions and matter in the universe are quantum. While experimentally we know that ordinary matter and strong and electroweak interactions are quantum, we don't have any experimental evidence that dark matter is quantum, or that gravity is quantum. There are theoretical arguments based on thought experiments (see, e.g., \cite{Eppley} and \cite{Page}, whose critiques appear, e.g., in \cite{Mattingly} and \cite{Hawkins}, respectively) why gravity should be quantum; however, currently there is no experimental evidence that gravity is in fact quantum (and, theoretical arguments aside, physics is an experimental science). The (perhaps technical, as opposed to conceptual) issue with keeping gravity classical is related to the question of how to couple it to quantum matter. The issues with semiclassical gravity (where gravity is coupled to the expectation value of the energy-momentum tensor for quantum fields) discussed in, e.g., \cite{Page} might be surmountable via a mechanism for wave function collapse \cite{Grossardt}. Coupling of classical gravity to quantum matter via stochastic interactions \cite{Oppenheim} is another avenue (albeit with its own open questions). One noteworthy aspect of the latter approach in the context of the initial conditions is that, perhaps the generation of random primordial density fluctuations can be attributable to the intrinsic stochastic interactions therein.     

\section{Concluding Remarks}

{}Let us conclude with some clarifying remarks. 

{}First, when discussing processes involving particle emission and absorption (e.g., photons), we venture outside the realm of non-relativistic quantum mechanics into relativistic quantum field theory (QFT) (e.g., quantum electrodynamics (QED)). In QFT we have the $CPT$ theorem (see, e.g., \cite{Blum} and references therein), which states that the theory is invariant under simultaneous $C$ (charge conjugation, or particle-antiparticle exchange), $P$ (spatial inversion, or parity transformation) and $T$ (time reversal) transformations. $CP$ is mostly conserved except in weak interactions involving kaons (or K-mesons), which contain a strange quark \cite{Kaon}. $CP$ violation implies violation of time-reversal in said interactions. However, it is unclear whether this violation is significant enough to account for the arrow of time on cosmic scales (cf. \cite{Vaccaro}). The same comment applies to $CP$ violation in the lepton sector (even if it explains the observed matter-antimatter asymmetry in the universe via leptogenesis; see, e.g., \cite{Davidson} and references therein). 

{}Speaking of QFT, typical theoretical objections to not quantizing gravity (see, e.g., \cite{Eppley} and \cite{Page}) essentially boil down to the following. The classical Einstein equation reads (in $c=1$ units):
\begin{equation}\label{EE}
 R_{\mu\nu} - \frac{1}{2}g_{\mu\nu}R = 8\pi G_N T_{\mu\nu}
\end{equation}  
Here $g_{\mu\nu}$ is the spacetime metric, $R_{\mu\nu}$ and $R$ are the tensor and scalar curvatures, and $T_{\mu\nu}$ is the energy-momentum tensor (for all matter and non-gravitational interactions). The latter becomes an operator ${\widehat T}_{\mu\nu}$ when matter and non-gravitational interactions are quantized. So, then the conundrum is how to make sense of (\ref{EE}) if its l.h.s. remains a number (rather than an operator). In semiclassical gravity one replaces $T_{\mu\nu}$ in the r.h.s. of (\ref{EE}) by the expectation value $\langle{\widehat T}_{\mu\nu}\rangle$ (see, e.g., \cite{Grossardt}), and the decades-old discussion ensues. 

{}However, before jumping the gun, one may wish to consider the following simple point. The issue is how to consistently couple QFT to classical gravity. However, does it even make sense to seek such ``consistent" coupling of QFT to classical gravity when QFT itself isn't self-consistent? Indeed, QED has a Landau pole \cite{Landau2}. In the electroweak theory the weak hypercharge group $U(1)_Y$ also has a Landau pole. An embedding of the strong and electroweak interactions into a grand unified theory (GUT) would avoid a Landau pole. However, there is no experimental evidence for GUTs, and their telltale sign, which is proton decay, hasn't been observed, albeit this doesn't rule out all GUTs, but does rule out some (for a recent discussion, see, e.g., \cite{proton} and references therein). One of the indirect ``hints" for a GUT is the observation that the electroweak and strong couplings approximately meet in the minimal supersymmetric standard model (MSSM), but not in its non-supersymmetric counterpart (see, e.g., \cite{Schwichtenberg} and references therein). However, there is still no experimental evidence for supersymmetry. (Non-supersymmetric GUTs can be constructed, but generally require additional matter fields at intermediate scales below the GUT scale, or large threshold corrections due to non-degenerate masses of superheavy fields ({\em Id}).) Furthermore, GUTs generally predict magnetic monopoles \cite{tHooft}, \cite{Polyakov}, which would be produced in the early universe \cite{Tye}, \cite{Einhorn}, and are then explained away by cosmic inflation \cite{Guth}. Inflation introduces its own issues (as discussed above), which then require additional ingredients, etc. And so on, and so forth, possibly {\em ad infinitum}.

{}The main purpose of the immediately preceding paragraph is to illustrate how solving one problem can turn into a self-fulfilling prophecy if one allows for tunnel vision to take over. The original problem was that of consistently coupling classical gravity to not-self-consistent QED. It's important to keep in mind that various computational techniques developed in physics (one of which is perturbative QED) aren't complete theories without limitations. They work well in certain regimes and/or certain computations, but aren't adequate in others. For instance, we don't compute the (zeroth-approximation) hydrogen atom spectrum using perturbative QED. Instead, we compute it using the non-relativistic Schr{\"o}dinger equation. To account for the fine structure of the hydrogen atom spectrum, we use the relativistic Dirac equation. To compute the Lamb shift and anomalous magnetic dipole moment of the electron, we use QED \cite{Bethe}, \cite{Schwinger}. These computational methods clearly have limitations. E.g., it's unclear how to make sense of QED to all orders in perturbation theory \cite{Dyson}, there's a Landau pole, etc. The lesson here is that overambitious, all-encompassing theories (apart from being very hard) often don't produce predictive computational tools. So, then the question is, pragmatically speaking, what's better, to spend decades on trying to couple gravity consistently to quantum matter and interactions (this includes both keeping gravity classical, and quantizing gravity, including via string theory (see, e.g., \cite{Becker})), or to find a computational method that would allow to make experimentally verifiable and falsifiable predictions? (Arguably, it's a matter of taste.)

{}Now, back to the arrow of time, it's all about the special initial conditions. The Trillion Dollar Question, once again, is how the special initial conditions in our universe arose. Cosmic inflation tries to answer some of the questions in this regard, at the expense of introducing others (see, e.g., \cite{Penrose2}, \cite{Earman}, \cite{Steinhardt}, and the discussion above). As discussed above, the question of special initial conditions is an open question for various reasons, including how the asymmetries we observe arose (and the inflationary scenario doesn't appear to answer this question). Therefore, the mystery of the apparent arrow of time, which is attributable to the special initial conditions, is a mystery of why, how, etc., these special initial conditions exist (albeit assuming randomness appears to suffice -- see above). Hopefully, this note, by stripping off various (arguably, unnecessary) complexities, which typically muddy the waters in discussions thereof, helps demystify the arrow of time, even if it doesn't answer the Trillion Dollar Question definitively.    

\subsection*{Acknowledgments}
{}I am grateful to my mother Mila Kakushadze for encouraging me to write up this note.

\end{document}